\documentclass[]{aa}

\usepackage{amsmath}
\usepackage{graphicx}
\usepackage{graphics}
\usepackage{epsfig}

\usepackage{txfonts}
\begin{document}

\title{Gaia-assisted discovery of a detached low-ionisation BAL quasar with very large ejection velocities}\titlerunning{A peculiar BAL quasar}

\author{
J.~P.~U.~Fynbo\inst{1,2},
P.~M\o ller\inst{3},
K.~E.~Heintz\inst{4},
J.~N.~Burchett\inst{5},
L.~Christensen\inst{6},
S.~J.~Geier\inst{7,8},
P. Jakobsson\inst{4},
J.-K.~Krogager\inst{9},
C.~Ledoux\inst{10},
B.~Milvang-Jensen\inst{1,2},
P.~Noterdaeme\inst{9},
J.~X.~Prochaska\inst{5},
T.~M.~Tripp\inst{11}
}
\institute{
Cosmic DAWN Center
\email{jfynbo@nbi.ku.dk}
\and
Niels Bohr Institute, University of Copenhagen, Lyngbyvej 2, 2100 Copenhagen \O, Denmark
\and
European Southern Observatory, Karl-Schwarzschildstrasse 2, D-85748 Garching, Germany
\and
Centre for Astrophysics and Cosmology, Science Institute, University of Iceland, Dunhagi 5, 107 Reykjav\'ik, Iceland
\and
University of California, Santa Cruz; 1156 High St., Santa Cruz, CA 95064, US
\and
DARK, Niels Bohr Institute, University of Copenhagen, Lyngbyvej 2, 2100 Copenhagen \O, Denmark
\and
Instituto de Astrof{\'i}sica de Canarias, V{\'i}a L{\'a}ctea, s/n, 38205, La Laguna, Tenerife, Spain
\and
Gran Telescopio Canarias (GRANTECAN), 38205 San Crist{\'o}bal de La Laguna, Tenerife, Spain
\and
Institut d'Astrophysique de Paris, CNRS-SU, UMR\,7095, 98bis bd Arago, 75014 Paris, France
\and
European Southern Observatory, Alonso de C\'ordova 3107, Vitacura, Casilla 19001, Santiago, Chile
\and
Department of Astronomy, University of Massachusetts, 710 North Pleasant Street, Amherst, MA 01003, US
}
\authorrunning{Fynbo et al.}

\date{Received 2019; accepted, 2019}

\abstract{
We report on the discovery of a peculiar Broad Absorption Line (BAL) quasar
identified in our Gaia-assisted survey of red quasars.
The systemic redshift of this quasar was difficult to establish due to the 
absence of conspicuous emission lines. 
Based on deep and broad BAL troughs (at least \ion{Si}{iv}, \ion{C}{iv}, and \ion{Al}{iii}),
a redshift of $z=2.41$ was established under the assumption that the
systemic redshift can be inferred from the red edge of the BAL troughs. 
However, we observe a weak and spatially-extended emission line
at 4450~\AA\ most likely due to Lyman-$\alpha$ emission, which implies a 
systemic redshift of $z=2.66$ if correctly identified. There is also evidence for the onset of Lyman-$\alpha$
forest absorption bluewards of 4450 \AA \ and evidence for H$\alpha$ emission in the
$K$-band consistent with a systemic redshift of $z=2.66$.
If this redshift is correct, the quasar is an extreme example of a detached low-ionisation BAL quasar.
The BAL lines must originate from material moving with very large velocities ranging from 22\,000 to 40\,000~km~s$^{-1}$.
To our knowledge, this is the first case of a systemic-redshift measurement based on extended Lyman-$\alpha$
emission for a BAL quasar, a method that should also be useful in cases of sufficiently distant BL Lac quasars without systemic-redshift information.
}

\maketitle

\section{Introduction}     
\label{sec:introduction}

The formation and growth of super-massive black holes in the centres of galaxies 
is a key element in the entire cosmic history of structure formation.
Understanding how super-massive black holes form also informs about
baryonic processes in galaxy formation, e.g., how gas is accreted.
The feedback processes from accreting super-massive black holes, in the form of active galactic nuclei or quasars, is a key element in regulating star formation \citep[see][for recent reviews]{Peterson2008,Fabian2012,
Netzer2015,Hickox2018}. More directly related to the object presented in this paper is the phenomenon of powerful outflows from quasars, in which material is seen being ejected from the central engine with very large velocities of 0.1-0.2\,$c$ \citep[see][for a review]{King2015}.

The spectral energy distributions of a large fraction of active galactic nuclei are 
remarkably similar, characterised by a very broad energy output extending over more 
decades in frequency space than for sources powered by nuclear fusion.
This broad spectral energy output forms the basis for a large range of selection methods for quasars, e.g.,
X-rays, ultraviolet excess, mm-emission, radio-emission, etc. \citep[see, e.g.,][and references therein]{Richards2006,Heintz2016}.
Photometric variability is also an important selection method for active galactic nuclei \citep[e.g.,][]{Ulrich1997,Schmidt2010}.

The object discussed in this short paper was discovered in a wider survey for quasars.
One of the main goals with this survey is to construct a more complete and (in terms of colour) more unbiased sample of quasars, partly based on astrometric data.
Specifically, the 2nd {\it Gaia} data-release \citep[{\it Gaia}-DR2;][]{Gaia2018} is used to select point-sources that are stationary, an indication that they may be extragalactic \citep{Heintz2015,Heintz2018b}.
This approach strongly reduces the contamination from stars during follow-up spectroscopy.

\citet[][see also \citealt{Krogager2016b}]{Heintz2018b} have in this manner built a catalogue of candidate quasars using a combination of astrometry from {\it Gaia}-DR2 and photometry from the optical Sloan Digital Sky Survey data release 12 \citep[SDSS-DR12,][]{Eisenstein2011}, the near-infrared UKIDSS \citep{Warren2007}, and the mid-infrared all-sky WISE mission \citep{Cutri13}.
For further details we refer to \citet{Heintz2018b} and \citet{Geier19}.
The object examined in this paper, designated GQ\,1309$+$2904 with equatorial coordinates of the target RA = 13:09:23.91, Dec = $+$29:04:51.7 (J\,2000.0), was observed as part of a larger spectroscopic campaign of candidate quasars from this survey.


We here present spectroscopic observations of the BAL quasar GQ\,1309$+$2904, and discuss its systemic redshift and the implied properties of the broad absorption lines.

\section{Observations and data reduction}    \label{sec:data}

The first spectroscopic observations of this candidate quasar were obtained with the OSIRIS instrument at the Gran
Telescopio Canarias (GTC). The candidate quasar was observed on January 5, 2019,
where three 500 sec integrations were obtained.
We used Grism R1000B and a 1.0 arcsec slit, providing a FWHM resolution of $\mathcal{R}=600$ and a spectral range of about 3750--7800~\AA.
Conditions were good with a seeing of about 0.9 arcsec.
We observed the target at low airmass, ranging from 1.02 to 1.04.
We secured additional spectroscopy (4$\times$800 sec) through a fast-track programme at the Nordic Optical Telescope on April 29, using the low-resolution spectrograph AlFOSC and Grism~20, covering the spectral range 6000--10100~\AA \ with a 1.0 arcsec wide slit providing a resolution of $\mathcal{R}\approx770$.
Conditions were again good with a seeing of about 1.0 arcsec and low airmass (below 1.03).
We subsequently observed the target in the near-IR, first with MOSFIRE at the Keck telescope in the $J$-band on June 5th 2019 for 1440 sec under $\sim0.6$ arcsec seeing.  We employed a 3-slitlet `longslit' setup resulting in a 21.6 arcsec long, 0.7 arcsec wide aperture with an ABAB dither pattern and, as recommended by the MOSFIRE team, 120s individual exposures.  
Finally, on June 12 2019, we observed the target with LIRIS mounted on the William Herschel Telescope (WHT).
We used the low-resolution HK grism and a 1.0 arcsec slit, covering the $H$ and $K$ bands at a resolution of 700. We obtained 15 exposures in ABAB sequences with a total integration time of 3600 sec.
These observations were obtained under good conditions with a seeing below 1 arcsec and at low airmass (1.2).

The spectroscopic data from the GTC, NOT, and WHT were reduced using standard procedures in 
IRAF\,\footnote{IRAF is distributed by the National Optical Astronomy
Observatory, which is operated by the Association of Universities for
Research in Astronomy (AURA) under a cooperative agreement with the
National Science Foundation.}. The spectra were flux calibrated using 
observations of the spectro-photometric standard stars Feige~110 and BD\,$+$26\,2606 
observed on the same nights. The MOSFIRE spectra were reduced using dedicated software.

We did not have a flux standard observation for the LIRIS data.
Instead, we observed the telluric A0V standard HIP 68209.
We used the spectrum of Vega to derive a relative sensitivity function and then scaled the fluxed spectra to the photometry in the $H$ and $K$ bands.

\section{Results}    \label{sec:results}

\subsection{Line identification}

\begin{figure*}[th]
\centering
\includegraphics[scale=1.00]{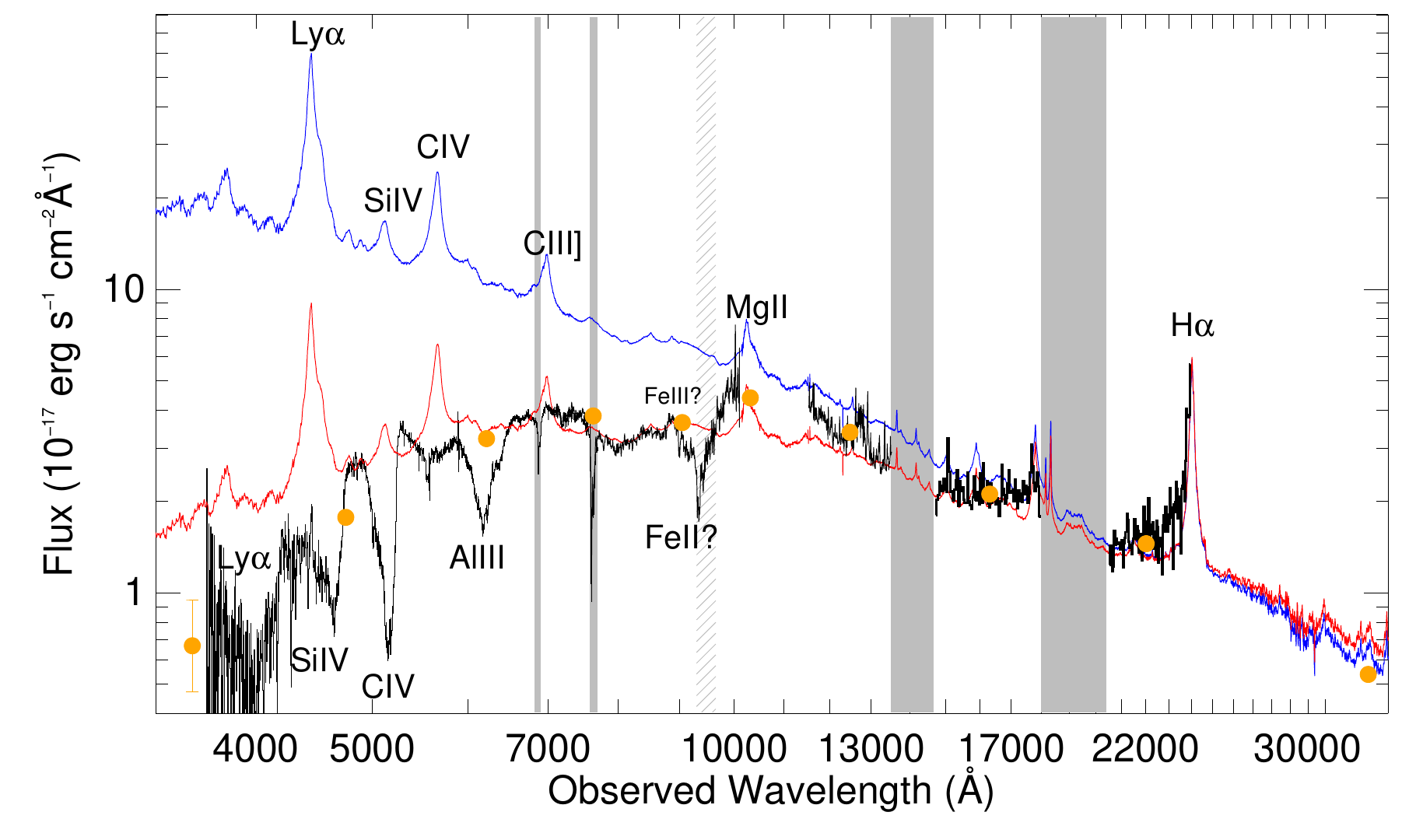}
\caption{The spectra of GQ\,1309$+$2904 from GTC, NOT, MOSFIRE, and LIRIS are shown together with $ugriz$, $YJHK_s$ and W1 photometry from the SDSS, UKIDSS, and WISE surveys (orange points).
The spectra are shown with logarithmic axes to better point out all important details.
The shaded areas around 6800~\AA\ and 7600~\AA\ indicate the locations of the telluric A and B absorption bands.
We have also shaded a region with many narrow telluric lines at 9300-9650~\AA, that may affect the appearance of, but cannot fully account for, the broad absorption lines seen there. 
The NOT spectrum has been smoothed with a five pixel-wide boxcar filter for illustration purposes.
In the MOSFIRE spectrum, we have interpolated across skyline residuals for illustration purposes.
Except for the $u$-band (where the error is 0.38 mag), the error-bars on the photometric data points are smaller than the plotting symbols.
Also overplotted are a template quasar spectrum from \citet{Selsing2016} redshifted to $z=2.66$ (in blue), and this same spectrum reddened by SMC-like extinction with $A_\mathrm{B}=0.55$ mag (in red).
}
\label{fig:spectrum}
\end{figure*}

In Fig.~\ref{fig:spectrum} our spectra of GQ\,1309$+$2904 are shown together with photometry from SDSS, UKIDSS and WISE.
The GTC spectrum shows a red shape and the presence of very broad and deep absorption features, the three clearest
being centred at around 6180~\AA, 5150~\AA, and 4640~\AA. 
It is obvious that the object is not a star. 
The source is also detected in all four WISE bands with colours consistent with the colours of quasars (Table~\ref{phot}, the $W1$-$W2$ colour was part of the selection criteria).
The source is not detected in the FIRST survey at radio frequencies \citep{Becker1995}, which is consistent with the properties of most BAL quasars \citep[][]{Morabito2019}.

The spectrum looks very different from normal type I quasars because of the absence of broad emission lines.
It also looks different from typical Broad Absorption Line (BAL) quasars in that the broad absorption lines are broader and with much less pronounced P-Cygni profiles than normally seen for BAL quasars \citep[e.g.,][]{Hall02,Gibson09}.
As an example, the BAL line at 5150~\AA\ has an observed equivalent width of more than 200 \AA.

\begin{table*}[!htbp]
\begin{tabular*}{1.0\textwidth}{@{\extracolsep{\fill}}c c c c c c c c c c c c c c c}
    \noalign{\smallskip} \hline \hline \noalign{\smallskip}
    \emph{u} & \emph{g} & \emph{r} & \emph{i} & \emph{z} & \emph{Y} & \emph{J} & \emph{H} & \emph{K$_s$} & W1 & W2 & W3 & W4 \\
    \hline
    22.84 & 21.13 & 19.88 & 19.25 & 18.92 & 17.80 & 17.39 & 16.83 & 16.13 & 15.43 & 14.51 & 10.95 & 8.52 \\
    \noalign{\smallskip} \hline \noalign{\smallskip}
\end{tabular*}
\caption{The optical and near-infrared magnitudes from SDSS \citep[AB magnitudes from][]{Alam15}, UKIDSS \citep[Vega magnitudes from][]{Warren2007}, and WISE \citep[Vega magnitudes from][]{Cutri13}.}
\label{phot}
\end{table*}

\begin{figure}[th]
\centering
\includegraphics[scale=0.47]{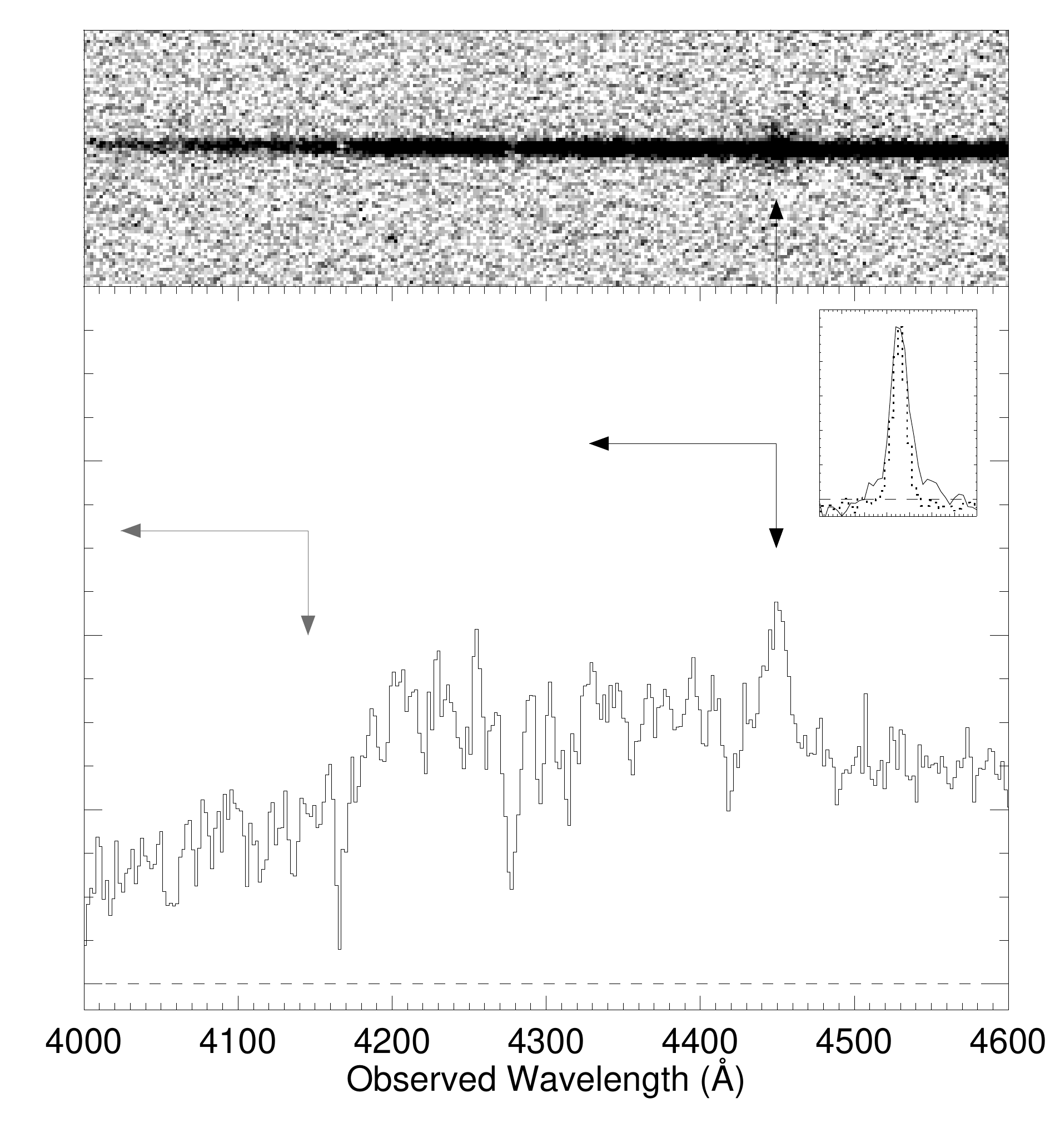}
\caption{A zoom-in on the blue part of the spectrum both in the 2d-spectrum (top panel) and 1d-spectrum (bottom panel).
The black arrows indicate the location of the spatially extended emission line, which we interpret as Lyman-$\alpha$ emission with a FWHM of $\approx1200$ km~s$^{-1}$, and the start of the Lyman-$\alpha$ forest at $z=2.66$.
Note that there are several narrow absorption lines shortwards of the alleged Lyman-$\alpha$ line, which would be consistent with the onset of the Lyman-$\alpha$ forest.
If the redshift was $z=2.41$, then the Lyman-$\alpha$ forest would only begin around 4145~\AA\ (marked with dark grey arrows).
The insert shows the spatial profile averaged over the wavelength range 4447--4455~\AA\ (solid line, covering the spatially extended emission line, which we interpret as Lyman-$\alpha$) compared to the continuum profile averaged over the wavelength range 4500--4600~\AA.
}
\label{fig:forest}
\end{figure}

\begin{figure}[th]
\centering
\includegraphics[scale=0.6]{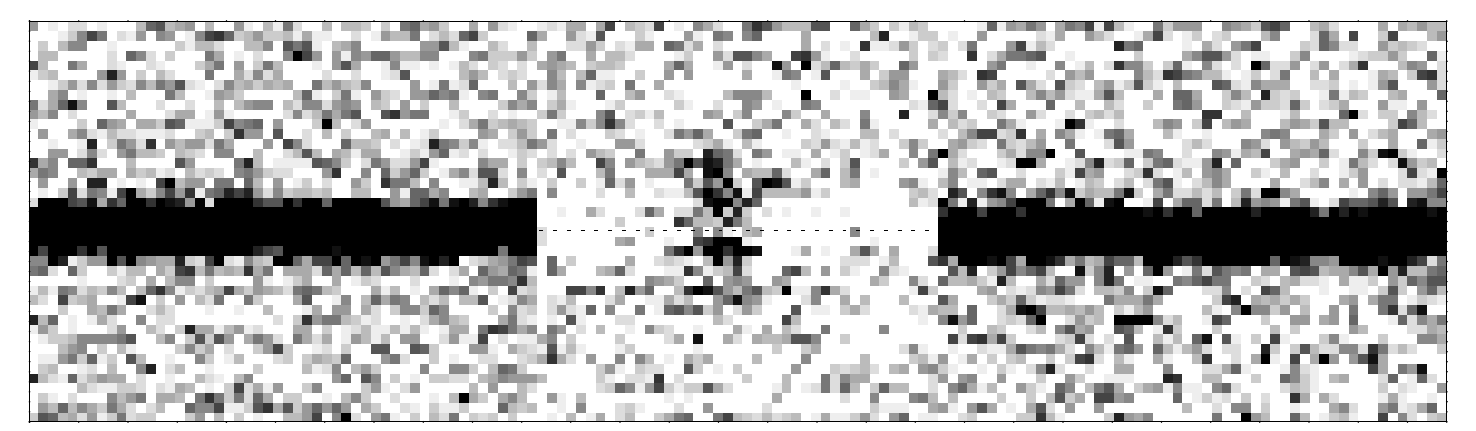}
\caption{We have here used spectral PSF subtraction to remove the light from the central unresolved emission component at 4450~\AA. We have subtracted a maximal point source contribution creating a hole in the middle of the spatial profile of the emission line. Even after this 
conservative procedure, we detect a clear, extended component on either side of the quasar trace
(marked with a dotted line).
These properties make the interpretation of this line as Lyman-$\alpha$ most likely compared to alternative explanations involving the superposition of two unrelated objects on the slit.}
\label{fig:extended}
\end{figure}

We first suspected that the three main troughs in the GTC spectrum of GQ\,1309$+$2904 were due to \ion{Si}{iv}, \ion{C}{iv}, and \ion{Al}{iii}, at a redshift of roughly $z=2.41$, based on the red-most edges of the troughs.
This would make GQ\,1309$+$2904 a low-ionisation BAL quasar \citep{Voit1993}.
However, upon closer inspection of the trace we identified a spatially extended emission line at 4450 \AA \ (see Figs. \ref{fig:forest} and \ref{fig:extended}).
The line is spatially asymmetric, extending more than 2 arcsec to one side of the trace and slightly less on the other side. 
This extended feature is present in all three individual GTC spectra and is hence not caused by a cosmic ray hitting close to the trace. To better show this
extended component, we have applied spectral point spread function (SPSF) subtraction following the procedure described in \citet{Moller2000}. We have conservatively scaled the SPSF such that the flux is zero at the centre of the trace. As seen in Fig.~\ref{fig:extended}, there is spatially extended emission
on either side of the trace.
The velocity width of the line is about 1200 km s$^{-1}$. 
It is natural to interpret this line as Lyman-$\alpha$, as extended Lyman-$\alpha$ emission is very commonly detected from luminous quasars, including BAL quasars \citep[e.g.,][]{Heckman91,Fynbo99,Moller00,Weidinger2005,Christensen06,Ginolfi2018,Cai2019}. The flux of the line is also in the range of those measured for extended Lyman-$\alpha$ emission around quasars at similar redshifts \citep{Cai2019}. This interpretation implies a significantly higher redshift of $z=2.66\pm0.01$.

\subsection{The NOT spectrum}
A systemic redshift of $z=2.66$ could explain the apparent flux-drop in the $z$-band compared to the $Y$-band, which at this redshift would be due to a \ion{Mg}{ii} emission line in the $Y$-filter, and possible BAL absorption due to \ion{Fe}{ii} in the $z$-band. The objective of the NOT spectrum was to check for this possibility.
In the NOT spectrum, there is indeed an additional deep absorption trough centred at around 9340~\AA, so the spectrum seems to be consistent with such an interpretation (compare, e.g., with the spectra of H\,1011$+$091 in \citet{Hartig1986}, HAQ\,1114+1330 in \citet{Krogager2015}, or eHAQ\,1514-0002 in \citet{Krogager2016b}). The NOT spectrum is rising towards the red extreme, which is consistent with a strong \ion{Mg}{ii} emission line in the $Y$-filter.

In the NOT spectrum, there is also evidence for an emission line consistent with \ion{Fe}{iii}\,$\lambda$2418 (also called \ion{Fe}{iii} UV47) at $z=2.66$, although it is difficult to infer with certainty if this feature is an emission line or a transmission peak between BAL troughs.
The \ion{Fe}{iii} UV47 emission line has previously been seen in BAL quasars \citep[][]{Laor1997,Vestergaard2001}. On the red wing of the telluric A absorption band, there is evidence for emission centred around 6980 \AA \ consistent with a weak \ion{C}{iii}] and/or \ion{Fe}{iii} UV34 at $z\approx2.66$ \citep[see][their fig.~5]{Vestergaard2001}.

\subsection{The near-IR spectra}

The MOSFIRE spectrum covering the $J$-band shows a flat continuum with a slope consistent with the photometry. There is a possible emission feature around 12600\AA, which we have not been able
to identify.
The LIRIS spectrum shows a flat spectrum in the $H$-band.
In the $K$-band, the flux is rising towards the red which is consistent with a strong H$\alpha$ emission line at $z=2.66$.
We can rule out the presence of a strong H$\alpha$ emission line at $z=2.41$.

\subsection{Lyman-$\alpha$ forest}
Consistent with the inferred redshift of $z=2.66$, we also observe evidence for the onset of the Lyman-$\alpha$ forest shortwards of 4450~\AA. If the redshift were $z=2.41$, the Lyman-$\alpha$ forest would begin at around 4145~\AA, but we do detect narrow absorption lines in the wavelength range 4145--4450~\AA\ (see Fig.~\ref{fig:forest}). No similarly narrow absorption lines are detected
redwards of 4450~\AA.

\begin{figure}[th]
\centering
\includegraphics[scale=0.75]{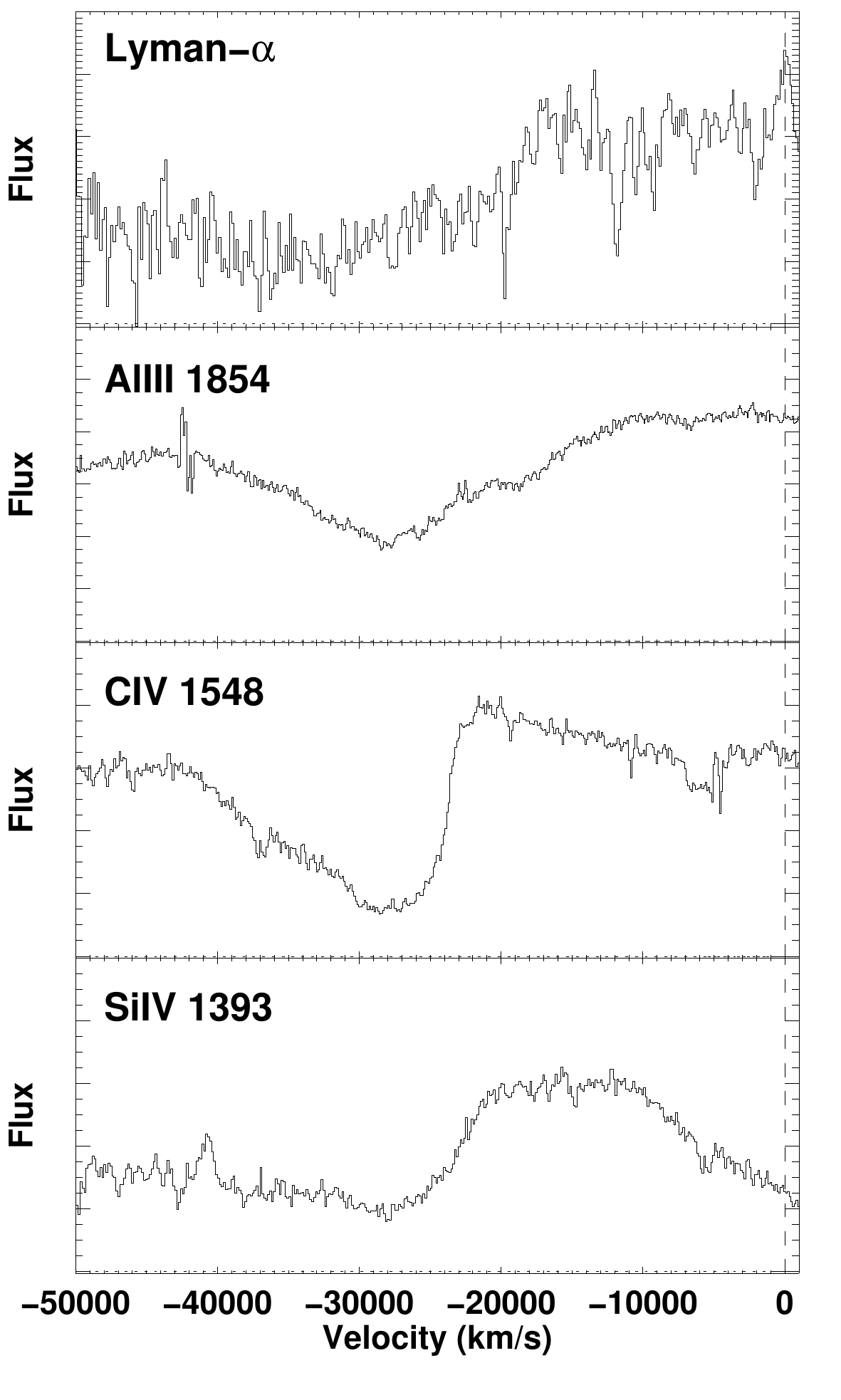}
\caption{The broad absorption lines plotted in velocity space shortwards of the proposed systemic redshift of $z=2.66$.
For the three doublets we have used the bluest components of the doublets, i.e., 1393~\AA, 1548~\AA, and 1854~\AA, for \ion{Si}{iv}, \ion{C}{iv}, and \ion{Al}{iii}, respectively.}
\label{fig:velocity}
\end{figure}

\subsection{Velocity structure of the BAL lines}

In Fig.~\ref{fig:velocity}, we plot the three main broad absorption troughs and Lyman-$\alpha$ under the assumption that the systemic redshift is $z=2.66$ and that the broad lines are due to \ion{Si}{iv}, \ion{C}{iv}, and \ion{Al}{iii}.
The velocities are calculated using eq. 1 in \citet{Foltz1986}. 
The inferred blue-shifts are extremely large extending up to terminal blue-shift velocities of about --40,000~km\,s$^{-1}$.
The \ion{C}{iv} trough has a quite well-defined red edge starting at a blue-shift of about --22,000~km\,s$^{-1}$. If we are right about the inferred systemic redshift of z=2.66, GQ\,1309$+$2904 is hence an extreme example of a detached BAL quasar.
For comparison, there are only 2--3 detached BAL quasars with minimum ejection velocities above 20,000~km\,s$^{-1}$ in the sample of more than 72 BALs studied by \citet{Lee1995}.
\citet{Benn2005} have also reported a quasar with a \ion{C}{iv} BAL detached by more than 20.000 km s$^{-1}$. 
In the spectrum of GQ\,1309$+$2904 there is also evidence for a weaker BAL component at a smaller velocity of about 6,000~km\,s$^{-1}$ seen both in \ion{C}{iv} and \ion{Si}{iv}.

\subsection{Photometric Variability}
\begin{figure}[ht!]
\centering
\includegraphics[scale=0.38]{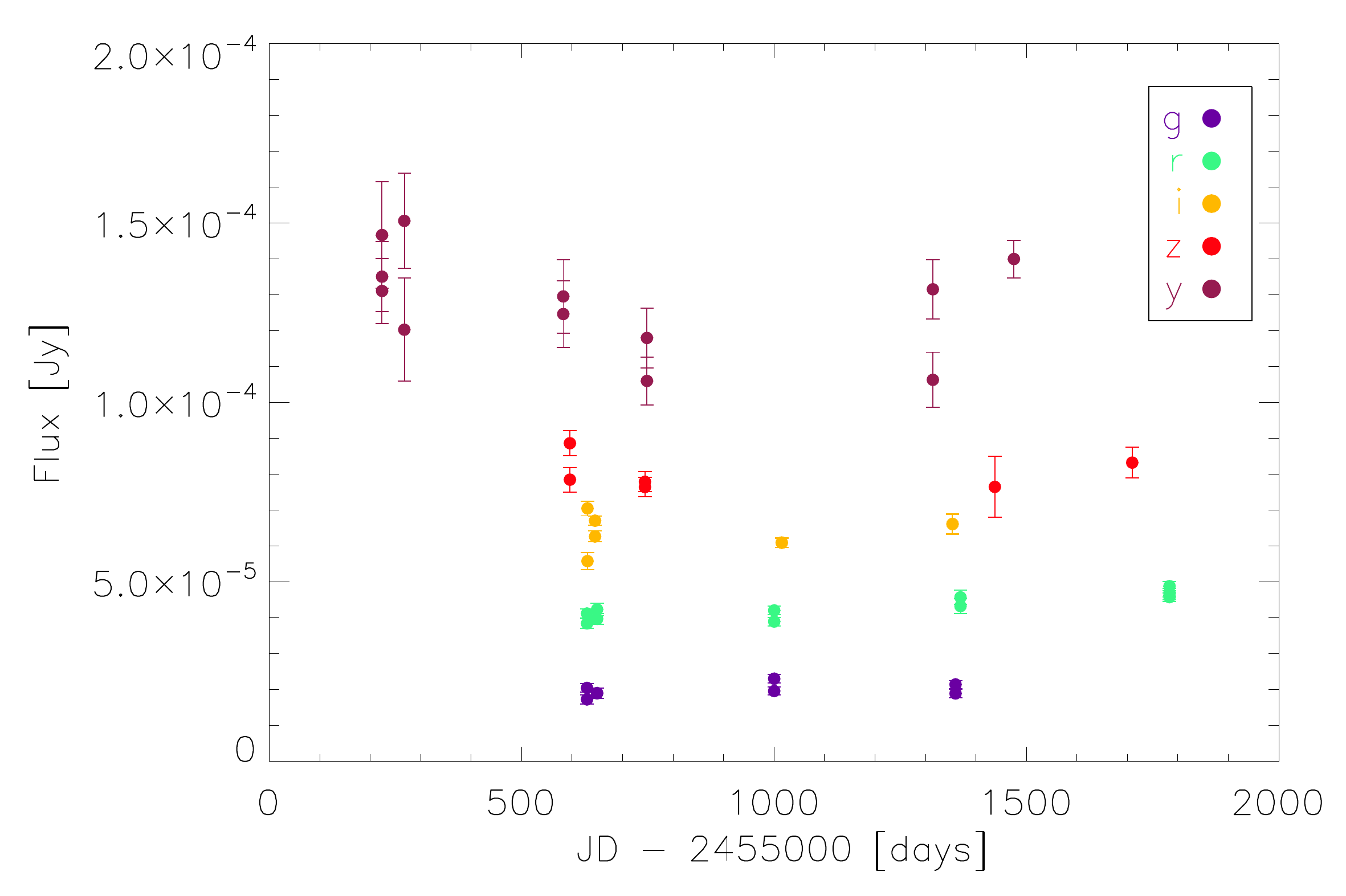}
\caption{Light curves of GQ\,1309$+$2904 from Pan-STARRS DR2 catalogue data in the $grizy$ filters demonstrate mean fractional variations of 5--7\% over the course of $2-3$ years.}
\label{fig:vari}
\end{figure}

The strength of the lines and the very large blue-shifts suggest that this source is likely to be variable \citep{Lundgren2007}.
The Pan-STARRS data release 2 catalogue contains data of the quasar field in the $grizy$ filters, obtained over a time span of $2-3$ years with $6-12$ measurement points per filter.
We extracted all photometric data, and corrected for atmospheric extinction.
Computing the mean fractional flux variation of the object following the definitions described in \citet{Peterson01}, which takes into account individual flux measurement uncertainties, we determine that in each filter the variability is $5-7$\%, which is within the range of normal variability for quasars, including BAL quasars  \citep{VandenBerk2004}.
Fig.~\ref{fig:vari} shows light curves from the five filters.

\subsection{Multiple sources on the slit?}

A possible, although unlikely, interpretation of our data is that there could be two sources at different redshifts superposed on the slit: an extended Lyman-$\alpha$ emitter at $z=2.66$ and a foreground BAL quasar at $z=2.41$.  
We have checked the existing imaging from SDSS, UKIDSS, and Pan-STARRS to search for any evidence for multiple sources along the line of sight.
However, in all bands the source is consistent with being a single point source and any additional source would have to be at least ten times fainter than the main component (this of course depends somewhat on the impact parameter).

Another possibility could be that the emission line at 4450 \AA \ is due to another emission line than Lyman-$\alpha$. Two arguments render this possibility unlikely. First, there is no evidence for other narrow emission lines next to 
the trace of any of our spectra, which we would expect if the line were for example due to [\ion{O}{ii}]. Second, the velocity width of the line is too large for 
such an interpretation. Very broad lines would belong to massive galaxies that should be detectable in existing imaging data.

\section{Discussion and conclusions} 
\label{sec:conc}

Normally, it is not difficult to estimate the systemic redshift of quasars
to within at least one percent, but the spectrum of GQ\,1309$+$2904 is very unusual compared to hundreds of BAL quasars we have observed in our surveys.
GQ\,1309$+$2904 is a low-ionisation BAL quasar with strong BAL trougs and
no apparent broad emission lines in the observed optical range.

It is not surprising that systems with large ejection velocities are discovered when searching for dust-obscured quasars \citep[e.g.,][]{SF1992}. 
If our redshift measurement based on the spatially extended line is correct the system
must be an extreme example of a detached low-ionisation BAL quasar. 
The onset of the main BAL trough is at a blue-shift of more than 22,000~km\,s$^{-1}$
and the trough extends to about 40,000~km\,s$^{-1}$. Ejection velocities as large as 0.1--0.2\,$c$ 
are not unheard of in BAL quasars \citep[e.g.,][]{Srianand01,King2015,Rogerson16,Hamann18},
but 40,000~km\,s$^{-1}$ is clearly in the high end of the distribution.

This interpretation depends critically on the accuracy of the systemic redshift.
The evidence for a systemic redshift of $z=2.66$ mainly consists of the presence of the spatially-extended 
and kinematically broad emission line at 4450~\AA, which we interpret as being due to Lyman-$\alpha$.
It would be very interesting to study this extended emission further using Integral Field Spectroscopy.
Supportive evidence for a systemic redshift of $z=2.66$ is {\it i)} the presence of narrow absorption lines consistent with the onset of Lyman-$\alpha$ forest absorption at larger redshifts than $z=2.41$, {\it ii)} possible weak emission lines consistent with \ion{Fe}{iii} and \ion{C}{iii}] at $z=2.66$, {\it iii)} the structure of the spectral energy distribution around the $z$- and $Y$-bands confirmed by the NOT spectrum is consistent with a BAL trough bluewards of a bright \ion{Mg}{ii} emission also consistent with $z=2.66$, and {\it iv)} the LIRIS spectrum that excludes strong H$\alpha$ emission at $z=2.41$ and shows tentative evidence for an H$\alpha$ line at $z=2.66$.

An important fact about the source is that it is very red.
To match the overall shape of the spectrum as well as the UKIDSS photometry extending to the $K_s$-band, we need to include an extinction of $A_B = 0.55$. 
BAL quasars constitute about 10-20\% of optically-selected quasars, but they are much more frequent among reddened quasars.
 \citet{Allen2011} also find an intrinsic BAL fraction in excess of 40\% based on analysis of the SDSS
quasar catalogue and a modelling of selection criteria. In our own surveys for reddened quasars we have found about half to be BAL quasars \citep{Krogager2015, Krogager2016b}.

Also, the absence of broad emission lines, which are normally observed in type-I quasars, is remarkable. However, weak or absent broad emission lines is not unusual for detached BAL quasars \citep{Hartig1986}. 

Interestingly, \citet{HL1998} and \citet{Lamy2004} have established an anti-correlation between the degree of polarisation and the magnitude of the detachment of the broad absorption lines.
This anti-correlation is particularly strong for low-ionization BALs like GQ\,1309$+$2904.
A way to test the validity of the systemic redshift proposed here is to measure the degree of polarisation.
This should be relatively easy given that the quasar is luminous.
If we are correct in establishing a systemic redshift of $z=2.66$ then the prediction from \citet{HL1998} and \citet{Lamy2004} is that the source should {\it not} be strongly polarised.

The presence of a large Lyman-$\alpha$ emitting halo around a galaxy hosting a BAL quasar is interesting in itself. It is an interesting question whether a quasar with so strong broad absorption lines is capable of releasing enough ionising radiation to illuminate the
surrounding circumgalactic medium as in the case of other quasars unless the BAL region is ``clumpy'' and not covering the full solid angle around the quasar.
Studies of the nearby BAL quasar MRK\,231 by \citet{Veilleux2013,Veilleux2016} have provided strong evidence that (at least some) BAL outflows indeed are patchy/clumpy, as might be required to explain the properties of the present BAL quasar. Moreover, in the models for BAL quasars discussed in, e.g., \citet{Lamy2004} and \citet{Veilleux2013,Veilleux2016} the viewing angle plays a central role in
determining whether we see a P-Cygni type BAL quasar or a detached BAL quasar like GQ\,1309$+$2904. However, the extended Lyman-$\alpha$ emission is most likely far less dependent on viewing angle given its large spatial extent (tens of kpc).

The existence of an extended Lyman-$\alpha$ emission halo also has consequences for the evolutionary state of the system. The lifetime of the quasar must be large enough to allow the setup of a large Lyman-$\alpha$ emitting halo, which sets a lower limit of order of 10$^5$ years simply based on the light-travel time across the halo.
Finally, to our knowledge, this is the first case of a systemic-redshift measurement based on extended Lyman-$\alpha$ emission of a BAL quasar.
This method should also be useful in cases of sufficiently distant BL Lac quasars with a priori unknown redshifts.

\begin{acknowledgements}
We thank the anonymous referee for a very helpful and constructive report.
We also thank Raghunathan Srianand, Yeimin Yi, Marianne Vestergaard, Sandra Raimundo, and Kimihiko Nakajima for helpful discussions. Based on observations made with the Gran Telescopio Canarias (GTC), the William Herschel Telescope (WHT), and with the Nordic Optical Telescope (NOT), installed in the Spanish Observatorio del Roque de los Muchachos of the Instituto de Astrofísica de Canarias, on the island of La Palma. Some of the data presented herein were obtained at the W. M. Keck Observatory, which is operated as a scientific partnership among the California Institute of Technology, the University of California and the National Aeronautics and Space Administration. The Observatory was made possible by the generous financial support of the W. M. Keck Foundation. 
The authors wish to recognise and acknowledge the very significant cultural role and reverence that the summit of Maunakea has always had within the indigenous Hawaiian community.  We are most fortunate to have the opportunity to conduct observations from this mountain. 
This work is partly based on data obtained through the UKIRT Infrared Deep Sky Survey.
The Cosmic Dawn Center is funded by the DNRF.
KEH and PJ acknowledge support by a Project Grant (162948--051) from The Icelandic Research Fund. 
LC is supported by the Independent Research Fund Denmark (DFF--4090-00079). JKK and PN acknowledge support from the French Agence Nationale de la Recherche under grant no ANR-17-CE31-0011-01 (project HIH2 -- PI: Noterdaeme). BMJ is supported in part by Independent Research Fund Denmark grant DFF - 7014-00017. JPUF thanks the Carlsberg Foundation for support.
\end{acknowledgements}

\bibliographystyle{aa}
 \newcommand{\noop}[1]{}

\object{GQ1309+2904}
\end{document}